\documentclass[conference]{IEEEtran}
\newcommand{\CLASSINPUTtoptextmargin}{2.4cm}
\newcommand{\CLASSINPUTbottomtextmargin}{2.6cm}

\IEEEoverridecommandlockouts

\usepackage{cite}
\usepackage{amsmath,amssymb,amsfonts}
\usepackage{graphicx}
\usepackage{algorithm}
\usepackage{algpseudocode}
\usepackage{textcomp}
\usepackage{xcolor}
\usepackage[font=footnotesize,skip=2pt]{caption}
\usepackage{subcaption}
\usepackage{hyperref}

\def\BibTeX{{\rm B\kern-.05em{\sc i\kern-.025em b}\kern-.08em
  T\kern-.1667em\lower.7ex\hbox{E}\kern-.125emX}}

\title{A Dual Belief-Driven Bayesian--Stackelberg Framework for Low-Complexity and Secure Near-Field ISAC Systems}

\author{
\IEEEauthorblockN{Mehzabien Iqbal, Ahmad Y. Javaid$^{*}$}
\IEEEauthorblockA{Electrical Engineering and Computer Science (EECS) Department\\
The University of Toledo, Toledo, OH, USA\\
mehzabien.iqbal@rockets.utoledo.edu, ahmad.javaid@utoledo.edu}
}

\begin{document}
\maketitle
\begin{abstract}
Ensuring robust security in near-field Integrated Sensing and Communication (ISAC) systems remains a critical challenge due to dynamic channel conditions, multi-eavesdropper threats, and the high computational burden of real-time optimization at mmWave and THz frequencies. To address these challenges, this paper introduces a novel Bayesian--Stackelberg framework that jointly optimizes sensing, beamforming, and communication. The dual-algorithm design integrates (i) Adaptive Hybrid Node Role-Switching between secure transmission and cooperative jamming (ii) Belief-Driven Sensing and Beamforming for confidence-based resource allocation. The proposed unified framework significantly improves robustness against attacks while preserving linear computational complexity. Simulation results across carrier frequencies ranging from 28 to 410 GHz demonstrate that the method achieves up to a 35\% increase in secrecy rates and a success rate exceeding 98\%, outperforming conventional communication systems with minimal runtime overhead. These findings underscore the scalability of belief-driven ISAC security solutions for low-complexity deployment in next-generation communications.
\end{abstract}

\section{Introduction}
\IEEEPARstart{S}{ixth}-generation (6G) wireless networks evolve from fifth-generation (5G) and Beyond-5G (B5G) paradigms to deliver ultra-high data rates, ultra-low latency, and intelligent connectivity \cite{chowdhury20206g, akyildiz20206g}. The utilization of millimeter-wave (mmWave) and terahertz (THz) frequency bands facilitates multi-gigabit communications. However, it also increases security vulnerabilities due to severe path loss, beam leakage, and the inherently broadcast nature of wireless channels\cite{10227884}. As AI-driven networks continue to advance, adversaries are expected to become increasingly adaptive, underscoring the need for integrated frameworks that jointly address performance and security.

\emph{Integrated Sensing and Communication (ISAC)} has thus emerged as a cornerstone technology for 6G \cite{10622357}, combining radar sensing and data communication within a unified framework to enhance spatial awareness and strengthen Physical Layer Security (PLS) \cite{9705498, 10770016}. Its dual-functional design enables joint waveform processing and hardware reuse, improving both spectral and energy efficiency \cite{10622473}. Consequently, ISAC has emerged as a central focus for secure, intelligent, and resource-efficient 6G systems. However, ISAC implementations in mmWave, sub-THz, and THz bands remain vulnerable to beam misalignment, reflections, and side-lobe leakage, which can expose confidential signals \cite{10622357}. The shared spectrum between sensing and communication further widens the attack surface \cite{10608156, 9761984}. Meanwhile, cryptographic schemes face scalability and energy constraints and may still leak information \cite{10227884}, while the fragility of mmWave/THz links to blockage and reconfiguration increases the risk of eavesdropping and jamming \cite{9405679, 9221122}. Recent approaches—such as Reconfigurable Intelligent Surfaces (RIS), near-field MIMO, and AI-based optimization improve secrecy and reliability \cite{10870062} but often depend on static or approximation-based models, limiting real-time adaptability. These challenges highlight the need for a unified, belief-driven, and dynamically reconfigurable framework that jointly addresses sensing, communication, and security trade-offs.

To overcome these challenges, game-theoretic and learning-based frameworks have recently gained prominence for modeling strategic interactions among legitimate users, and eavesdroppers \cite{10543060}. However, most existing studies focus on individual objectives, such as beamforming or power control, and often neglect the joint interplay between minimizing computational complexity, robust security, and dynamic resource adaptation \cite{10812728}, which are critical for robust ISAC operation in dense and heterogeneous 6G networks.

Motivated by these challenges, this work proposes a Bayesian–Stackelberg game-theoretic framework for secure, low-complexity, and adaptive near-field MIMO-ISAC systems. The key contributions are summarized below:
\begin{itemize}
\item \textbf{Adaptive Hybrid Node Role-Switching:} We introduce a dual-mode Hybrid Node (HN) architecture that allows each user to dynamically alternate between secure transmission and cooperative jamming based on instantaneous secrecy thresholds, thereby enhancing physical-layer security with minimal coordination overhead.
\item \textbf{Bayesian–Stackelberg Game-Theoretic Control:}  We propose a Bayesian–Stackelberg belief-driven controller within the ISAC framework to model strategic interactions between the ISAC base station (leader) and hybrid nodes (followers), achieving adaptive and efficient security optimization under uncertainty.
\item \textbf{Adaptive Belief-Driven Sensing and Beamforming:} We develop a dual-belief, low-complexity per-slot algorithm that integrates Bayesian prediction with confidence-guided beam design to jointly allocate sensing and communication resources. By updating probability distributions and kernel parameters in real time, the method enables robust beam adaptation against multi-eavesdropper and dynamic channel conditions.
\end{itemize}
\section{System Model and Problem Formulation}
\label{Sysmodel_problem}
\subsection{System Architecture}
\label{model}
\begin{figure}[h!]
\centering
\includegraphics[width=1.0\columnwidth]{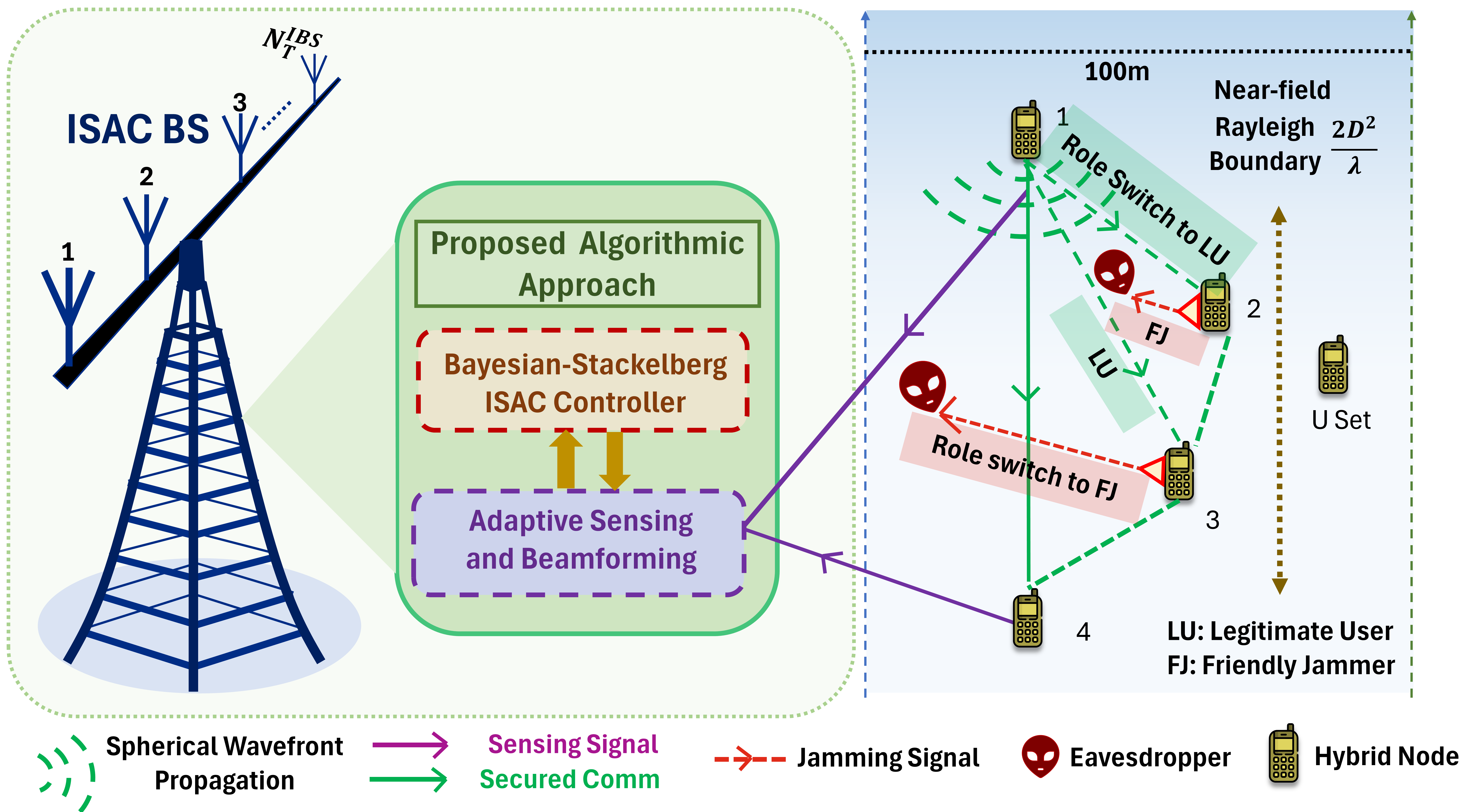}
\caption{Dual belief driven Bayesian-Stackelberg System Model for Near-Field ISAC Systems}
\label{Sysmodelfigl}
\end{figure}
\subsubsection{Network Model}
\label{net_model}
We consider ISAC based network model operating across the mmWave, sub-THz, and THz bands, comprising a single ISAC base station (IBS) considering hybrid MIMO-based ISAC transceiver, a set of $U$ hybrid nodes (HNs), and potential adversarial entities. Additionally, the proposed framework operates under a Time Division Duplex (TDD) protocol to facilitate channel reciprocity and enable joint communication–sensing optimization. Therefore, in our system, the ISAC BS serves $U$ HNs located within a maximum coverage radius of $100$\,m, operating at a carrier frequency $f_c$ with wavelength $\lambda = c/f_c$. The entire service region lies within the radiative near-field since the Rayleigh boundary, defined as $R_{\mathrm{Rayleigh}} \triangleq \frac{2D^2}{\lambda}, \label{eq:rayleigh}$ \cite{daei2025near}. Consequently, all HNs within $100\,m$ are modeled under spherical wavefront propagation conditions. The ISAC BS operates with an $L{:}M$ downlink-to-uplink (DL/UL) slot ratio. Each HN can dynamically alternate between two roles per TDD frame: (i) acting as a egitimate receiver in the DL phase to receive confidential data, or (ii) functioning as a cooperative friendly jammer in the UL phase to transmit artificial noise (AN) or jamming signals toward potential eavesdroppers. Furthermore, sensing functionality is co-located at the ISAC BS and reuses DL resources, achieving a unified secured near-field ISAC operation.

\subsubsection{Near Field Antenna Arrays and Geometry, and RF Chains}
\label{antennaandRF}
In the proposed ISAC system model (\ref{model}), the base station (BS) employs a large-scale hybrid MIMO array with $N_T^{\mathrm{IBS}}$ antennas and $N_{\mathrm{RF}}$ RF chains in a partially connected structure. The overall precoder is \begin{equation}
    \mathbf{F} = \mathbf{F}_{\mathrm{RF}} \mathbf{F}_{\mathrm{BB}} 
    \in \mathbb{C}^{N_T^{\mathrm{IBS}} \times K}
    \label{eq:precoding},
\end{equation}
where $\mathbf{F}_{\mathrm{RF}} \in \mathbb{C}^{N_T^{\mathrm{IBS}} \times N_{\mathrm{RF}}}$ represents the analog RF precoder implemented through phase shifters (block-diagonal under partial connectivity), and $\mathbf{F}_{\mathrm{BB}} \in \mathbb{C}^{N_{\mathrm{RF}} \times K}$ denotes the digital baseband pre-coder for multi-user interference mitigation \cite{10897852}. Each RF chain drives a subarray of size $N_{\mathrm{SA}} = N_T^{\mathrm{IBS}} / N_{\mathrm{RF}}$. Both BS and Hybrid Nodes (HNs) employ uniform linear arrays (ULAs) with half-wavelength spacing ($d = 0.5\lambda$), ensuring grating-lobe-free beam control at THz frequencies.

For near-field focusing, the $n$-th BS antenna element is located at $\mathbf{r}n = [0,,nd,,0]^T$, and the focal point $\mathbf{p} = [x_p,,y_p,,z_p]^T$ yields the propagation distance
\begin{equation}
r_n(\mathbf{p}) = \sqrt{x_p^2 + (y_p - nd)^2 + z_p^2},
\end{equation}
with phase $\phi_n(\mathbf{p}) = k_0 r_n(\mathbf{p})$. The near-field steering vector,
\begin{equation}
\mathbf{a}_{\mathrm{NF}}(\mathbf{p}) = \frac{1}{\sqrt{N_T^{\mathrm{IBS}}}}
\begin{bmatrix}
e^{-jk_0 r_{-\frac{(N_T^{\mathrm{IBS}}-1)}{2}}(\mathbf{p})} \\
\vdots \\
e^{-jk_0 r_{\frac{(N_T^{\mathrm{IBS}}-1)}{2}}(\mathbf{p})}
\end{bmatrix},
\end{equation}
which depends jointly on range and angle, enabling precise beam focusing for near-field MIMO operation. HNs adopt similar steering formulations scaled to their apertures.

\subsubsection{Near-Field MIMO Channel}
\label{channel_model}
We adopt a clustered narrowband Saleh–Valenzuela (S–V) channel model extended to account for spherical wavefront propagation \cite{10897852}. Accordingly, the ISAC BS–to–HN channel for a legitimate user, $lu$ is characterized by $L$ dominant Multipath Components (MPCs) as
\begin{equation}
    \mathbf{H}_{lu} = 10^{-{\mathrm{PL}}_{lu}/20}
    \sum_{\ell=1}^{L} \alpha_{\ell,lu}\,
    \mathbf{a}_{\mathrm{HN,rx}}\!\big(\mathbf{p}^{\mathrm{AoA}}_{\ell,lu}\big)\,
    \mathbf{a}_{\mathrm{BS,tx}}^{H}\!\big(\mathbf{p}^{\mathrm{AoD}}_{\ell,lu}\big),
\end{equation}
where $\alpha_{\ell,lu}\!\sim\!\mathcal{CN}(0,1)$ denotes the complex path gain, and $\mathbf{a}_{\mathrm{BS,tx}}(\cdot)$ and $\mathbf{a}_{\mathrm{HN,rx}}(\cdot)$ are the near-field steering vectors.Similarly, for inter-HN channels $\mathbf{H}_{lu\leftarrow v}$, an identical structure applies with corresponding transmit and receive steering vectors. Moreover, at higher frequencies, large-scale attenuation incorporates both free-space spreading and molecular absorption, modeled as
\begin{equation}
    \mathrm{PL}(d,f_c) = 20\log_{10}\!\Big(\tfrac{4\pi f_c d}{c}\Big)
    + \kappa(f_c)\, d + X_\sigma,
\end{equation}
where $\kappa(f_c)$ represents the frequency-dependent absorption coefficient and $X_\sigma$ captures log-normal shadowing effects. Given $d \le 100$\,m and the Rayleigh boundary condition \eqref{eq:rayleigh}, the LoS component is typically dominant ($L$ small), and hence spherical-phase curvature must be preserved.Finally, to capture estimation uncertainty, imperfect CSI is modeled using a bounded error formulation \cite{10897852}:
\begin{equation}
    \mathbf{H} = \widehat{\mathbf{H}} + \mathbf{E}, 
    \qquad \|\mathbf{E}\|_F^2 \le \epsilon,
\end{equation} where $\widehat{\mathbf{H}}$ is the estimated channel and $\mathbf{E}$ represents the bounded estimation error, with separate error limits defined for legitimate and eavesdropping links.

\subsubsection{Signal Model, Secrecy Metrics, and Sensing Model}

Under hybrid precoding and linear combining, the received signal at the $u$-th hybrid node (HN) from the ISAC BS can be expressed as
\begin{equation}
\mathbf{y}_u = \mathbf{W}_u^{H} \left( \mathbf{H}_{lu}\mathbf{F}_{\mathrm{RF}}\mathbf{F}_{\mathrm{BB}}\mathbf{s} + \sum_{j\in\mathcal{J}} \mathbf{H}_{lu\leftarrow j}\mathbf{g}_j x_j + \mathbf{n}_u \right),
\end{equation}
where $\mathbf{s}\!\in\!\mathbb{C}^{K}$ is the transmitted data vector, $\mathcal{J}$ is the set of cooperative jammers, $\mathbf{g}_j$ represents the jamming beam, and $\mathbf{n}_u\!\sim\!\mathcal{CN}(\mathbf{0}, N_0\mathbf{I})$ denotes thermal noise. The instantaneous SINRs at the $U$-th legitimate HN and at the eavesdropper are respectively given by
\begin{align}
    \mathrm{SINR}_{lu,u} &=
    \frac{P_{\mathrm{TX}}\!\left|\mathbf{W}_u^{H}\mathbf{H}_{lu}\mathbf{f}_u\right|^2}
         {\sum_{k\neq u} P_{\mathrm{TX}}\!\left|\mathbf{W}_u^{H}\mathbf{H}_{lu}\mathbf{f}_k\right|^2
          + I_u + N_0}, \label{eq:SINR_u}\\
    \mathrm{SINR}_{e,u} &=
    \frac{P_{\mathrm{TX}}\!\left|\mathbf{W}_e^{H}\mathbf{H}_e\mathbf{f}_u\right|^2}
         {\sum_{k\neq u} P_{\mathrm{TX}}\!\left|\mathbf{W}_e^{H}\mathbf{H}_e\mathbf{f}_k\right|^2
          + I_e + N_0}, \label{eq:SINR_e}
\end{align}
where $\mathbf{H}_{lu}$ and $\mathbf{f}_k$ embed the near-field beamforming effects. Here, $I_u$ and $I_e$ denote the interference observed at the legitimate HN and eavesdropper, respectively, while $N_0$ represents the receiver noise power. Following this, the achievable secrecy rate for the $u$-th legitimate link is expressed as
\begin{equation}
    R_s^{(lu)} =
    \Big[
        \log_2(1+\mathrm{SINR}_{lu}) -
        \log_2(1+\mathrm{SINR}_{e,lu})
    \Big]^+,
    \label{eq:secrecyrateHN}
\end{equation}
and, for the worst-case eavesdropping scenario with $\mathrm{SINR}_e^{\max}$,
\begin{equation}
    R_s =
    \Big[
        \log_2(1+\mathrm{SINR}_{lu}) -
        \log_2(1+\mathrm{SINR}_e^{\max})
    \Big]^+,
    \label{eq:secrecyrate}
\end{equation}

For sensing, a power-normalized near-field focusing beam $\mathbf{w}_{\mathrm{S}}(\mathbf{p})$ directed toward a focal point $\mathbf{p}$ produces the response
where $\mathbf{a}_{\mathrm{IBS}}(\mathbf{p})$ denotes the ISAC BS near-field steering vector. To flexibly control beamwidth and sidelobe levels, the sensing beam is adaptively shaped as
\begin{equation}
    \mathbf{w}_{\mathrm{S}}(\tau,\mathbf{p}) =
    \Big[(1-\tau)\mathbf{1}_{N_T^{\mathrm{IBS}}} + \tau\,\mathbf{h}_{\mathrm{Hamming}}\Big]
    \odot \mathbf{a}_{\mathrm{IBS}}(\mathbf{p}), \tau\!\in\![0,1],
\end{equation}
where $\tau$ regulates tapering: $\tau=0$ corresponds to a broad field-of-view beam, while $\tau=1$ yields a narrow, high-gain Hamming-tapered focus \cite{10841413}. This range-adaptive shaping preserves co-focal alignment between sensing and communication beams under near-field conditions. 
\subsection{Problem Formulation}
\label{problem}
In ISAC-enabled wireless networks, ensuring physical-layer security is essential against both passive and active eavesdroppers sharing the same spectrum. The goal is to ensure that each legitimate user achieves a secrecy rate above the threshold $R_{\mathrm{th}}$ with high reliability. Thus, for this problem, core requirement can be expressed as a probabilistic constraint:
\begin{equation}
\Pr\{R_s(u) < R_{\mathrm{th}}\} \leq \varepsilon, \qquad \forall u \in \mathcal{A},
\end{equation}
where $\varepsilon$ denotes the maximum allowable secrecy-outage probability.
Again, the instantaneous secrecy-outage indicator be
\begin{equation}
\begin{aligned}
\mathbf{I}_{\mathrm{out}}(u) &= 
\begin{cases}
1, & R_s(u) < R_{\mathrm{th}}, \\
0, & \text{otherwise},
\end{cases} \\
P_{\mathrm{out}}^{\mathrm{sys}} &= \frac{1}{|\mathcal{A}|} \sum_{u \in \mathcal{A}} \mathbf{I}_{\mathrm{out}}(u)
\end{aligned}
\end{equation} representing the fraction of HNs failing to meet secrecy requirements. 
Each HN operates in a dual-functional mode, switching between legitimate communication and cooperative jamming based on its instantaneous secrecy state. Therefore, for the optimal role switching the friendly jamming (FJ) optimization problem is formulated as:
\begin{align}
\min_{\{\mathsf{Jam}(u)\},\, \mathbf{g}} \quad 
& P_{\mathrm{out}}^{\mathrm{sys}}(\mathsf{Jam}, \mathbf{g}), 
\label{eq:FJ_obj}\\[3pt]
\text{s.t.} \quad 
& \mathsf{Jam}(u) \in \{0,1\}, \quad \forall u \in \mathcal{A}, 
\label{eq:FJ_binary}\\[3pt]
& P_{\mathrm{FJ}} = n_J P_{u}^{\max} \leq P_{\mathrm{FJ}}^{\max}, 
\label{eq:FJ_power}\\[3pt]
& P_{\mathrm{leak}}^{(u)} \le \rho_{\mathrm{leak}} P_{\mathrm{lu}}^{\max}, 
\quad \forall u \in \mathcal{A}, 
\label{eq:FJ_leak}
\end{align}
where $\mathbf{g}$ represents the collective jamming beamforming vector. The objective in \eqref{eq:FJ_obj} minimizes the overall secrecy-outage probability, while constraints \eqref{eq:FJ_power}--\eqref{eq:FJ_leak} ensure power efficiency and control interference toward legitimate users. The goal is to jointly optimize jamming activation and beamforming for resilient, low-complexity secrecy under imperfect CSI and dynamic sensing conditions.
\section{Proposed Unified Framework}
\label{Proposedframework}
\subsection{Bayesian–Stackelberg Framework and Adaptive Control}
\label{Stack_bayes_com_sec_power}
To solve the problem described in \ref{problem} and to enable hierarchical and adaptive physical-layer security under channel and sensing uncertainty, we proposed a framework utlilizing \emph{Bayesian–Stackelberg game}~\cite{11172669,10608156,10543060}, 
where the ISAC Base Station acts as the leader, and Hybrid Nodes serve as followers. Each time slot $t$ involves Bayesian belief prediction, leader optimization, follower response, and meta-adaptation. The goal of this hierarchical game mechanism is to jointly integrate belief-driven Stackelberg optimization, adaptive HN role-switching, to achieve robust secrecy with low latency and computational efficiency under imperfect CSI and dynamic threat conditions. The decision sequence is summarized in Algorithm~\ref{alg:isac_BayStack}.
Moreover, to satisfy problems discussed in \ref{problem}, we denoted $\alpha$, $\beta$, and $\gamma$,
$\alpha + \beta + \gamma = 1,$
where $\alpha$ governs secure communication power, $\beta$ controls jamming and interference mitigation, and $\gamma$ determines the sensing intensity.

\subsubsection{Bayesian Belief Prediction and Update}
\label{beliefstate_SB}
At the beginning of slot $t$, the leader maintains a discrete angular belief 
$\mathbf{p}_t \in \Delta^{N_\phi-1}$ over possible eavesdropper directions. Threfore, the belief evolves as
\begin{align}
\mathbf{p}^{\mathrm{pred}}_t &= \mathbf{K}(\sigma_t)*\mathbf{p}_{t-1}, &
\mathbf{p}_t &\propto \mathbf{L}_t \odot \mathbf{p}^{\mathrm{pred}}_t,
\end{align}
where $\mathbf{K}(\sigma_t)$ denotes a Gaussian-like transition kernel—a discrete, symmetric, normalized kernel that spreads prior probability across nearby azimuth bins to model smooth angular drift, and $\mathbf{L}_t$ represents pseudo-likelihoods from tapered sensing beams steered toward the Maximum a Posteriori (MAP) peaks. 
Additionally, residual uncertainty is quantified by the Shannon entropy $H(\mathbf{p}_t)$, 
which governs the entropy-aware sensing fraction
\begin{equation}
\gamma_t = \Gamma(H(\mathbf{p}_t)) \in [\gamma_{\min},\,\gamma_{\max}],
\label{entropyaware}
\end{equation} ensuring stronger sensing when uncertainty is high.
\subsubsection{Leader Optimization and Policy Announcement}
Given the updated posterior $\mathbf{p}_t$ \eqref{entropyaware}, the leader (ISAC BS) predicts each HN’s expected secrecy rate $\widehat{R}^{(u)}(t)$ and announces a compact control policy
\[
\theta_t = \{\gamma_t,\,\mathcal{B}_t,\,\rho_{\mathrm{leak}}\},
\]
where $\gamma_t$ is the entropy-driven sensing weight, $\mathcal{B}_t$ denotes 
protected legitimate-user angles (beamforming notches), and $\rho_{\mathrm{leak}}$ is the leakage cap limiting interference to legitimate receivers. This decision anticipates the followers’ best responses and establishes the leader’s Stackelberg equilibrium action.

\subsubsection{ Follower Role Selection and Cooperative Response}
Upon receiving $\theta_t$, each HN applies a local best-response rule following the equation described in problem formulation, \eqref{eq:FJ_obj}:
where $R_{\mathrm{th}}$ is the target secrecy threshold from \ref{problem}. Thus, nodes with insufficient predicted secrecy act as friendly jammers (FJs), while others continue secure transmission. 
\subsubsection{Dual Updates and Meta-Adaptation}
After observing the resulting secrecy rate $R_s^{(u)}(t)$, the leader computes key performance indicators (KPIs): 
average secrecy $\overline{R}_s$, secrecy-outage probability $P_{\mathrm{out}}$, 
and posterior entropy $H(\mathbf{p}_t)$. 
The dual variables $\lambda_O$ and $\lambda_H$ enforce secrecy through
\begin{align}
\lambda_O &\leftarrow [\lambda_O + \mu_O(R_{\mathrm{th}}-\overline{R}_s)]_+,\\
\lambda_H &\leftarrow [\lambda_H + \mu_H(H(\mathbf{p}_t)-H^\star)]_+,
\end{align}
where $\mu_O$ and $\mu_H$ are learning rates, and $H^\star$ is the desired entropy bound. 
The kernel width $\sigma_t$ is then adapted as
\begin{equation}
\sigma_{t+1}\!\gets\!\operatorname{clip}\!\big(\sigma_0[1{+}k_{\rm ex}\mathbf{1}\{\overline{R}_s<R_{\mathrm{th}}\}
{-}k_{\rm ey}\mathbf{1}\{\overline{R}_s\ge R_{\mathrm{th}}\}]\big),
\end{equation}balancing exploration (wider sensing under poor secrecy) and exploitation (focused sensing under stable secrecy).
\subsubsection{Utility and Low-Complexity Objective}
The leader maximizes a low-complexity, belief-driven utility function:
\begin{equation}
U_t = \omega_R \min_{u\in\mathcal{A}_t} R_s^{(u)}(t)
- \omega_J P_{\mathrm{FJ}}(t)
- \lambda_H[H(\mathbf{p}_t)-H^\star]_+,
\label{eq:leader-utility-bayesian}
\end{equation}
where, $\omega_R$ and $\omega_J$ weight secrecy improvement and cooperative-jamming cost, respectively, 
and $P_{\mathrm{FJ}}(t)$ is the total FJ transmit power. 
This formulation ensures rapid convergence and computational efficiency, enabling real-time, belief-aware adaptation in near-field MIMO-ISAC environments.
\begin{algorithm}[t]
\caption{Bayesian–Stackelberg ISAC Controller (Per Slot)}
\label{alg:isac_BayStack}
\begin{algorithmic}[1]
\Require Prior belief $\mathbf{p}_{t-1}$, kernel width $\sigma_t$, duals $(\lambda_O,\lambda_H)$, secrecy threshold $R_{\mathrm{th}}$
\Ensure Posterior $\mathbf{p}_t$, kernel width $\sigma_{t+1}$, jammer set $\{\mathsf{Jam}_t(u)\}$, KPIs $(\overline{R}_s,P_{\mathrm{out}},H)$

\State \textbf{Leader (ISAC BS): Bayesian update}
\State Predict belief: $\mathbf{p}^{\mathrm{pred}}_t \!\gets\! \mathbf{K}(\sigma_t)*\mathbf{p}_{t-1}$
\State Steer tapered sensing beams to MAP peaks and update posterior 
$\mathbf{p}_t \!\propto\! \mathbf{p}^{\mathrm{pred}}_t \odot z(\phi)^{k_{\mathrm{eff}}}$;
compute entropy $H(\mathbf{p}_t)$
\State Adapt sensing weight: $\gamma_t\!\gets\!\Gamma(H(\mathbf{p}_t))$
\State \textbf{Leader: Stackelberg decision}
\State For each $u$, predict $\widehat{R}^{(u)}(t)$; 
announce policy $\theta_t=\{\gamma_t,\mathcal{B}_t,\rho_{\mathrm{leak}}\}$ to followers
\State \textbf{Followers (HNs): Best response}
\State $\mathsf{Jam}_t(u)\!\gets\!\mathbf{1}\{\widehat{R}^{(u)}(t)<R_{\mathrm{th}}\}$;
construct friendly-jamming beams aligned with $\mathbf{p}_t$, notched at $\mathcal{B}_t$
\State \textbf{Leader: Evaluation \& adaptation}
\State Compute $R_s^{(u)}$, $\overline{R}_s$, $P_{\mathrm{out}}$, and $H$;
update duals
$\lambda_O\!\leftarrow\![\lambda_O+\mu_O(R_{\mathrm{th}}-\overline{R}_s)]_+$,
$\lambda_H\!\leftarrow\![\lambda_H+\mu_H(H-H^\star)]_+$
\State Update kernel width: 
$\sigma_{t+1}\!\gets\!\operatorname{clip}\!\big(\sigma_0[1{+}k_{\rm ex}\mathbf{1}\{\overline{R}_s<R_{\mathrm{th}}\}{-}k_{\rm ey}\mathbf{1}\{\overline{R}_s\ge R_{\mathrm{th}}\}]\big)$
\end{algorithmic}
\end{algorithm}
 
\subsection{ISAC Adaptive Sensing and Beamforming Framework}
\label{ISACadaptive}
Building upon the Bayesian–Stackelberg controller (Algorithm~\ref{alg:isac_BayStack}), 
the Adaptive Sensing and Beamforming module (Algorithm~\ref{alg:adaptive}) 
operates as a subordinate routine that refines the ISAC BS’s belief, adjusts beam directions for the subsequent Stackelberg decision. While Algorithm~\ref{alg:isac_BayStack} governs the leader–follower control, role switching, and secrecy utility optimization, Algorithm~\ref{alg:adaptive} performs intra-slot adaptation providing the leader with an updated posterior $\mathbf{p}_t$, sensing fraction $\gamma_t$, and optimized beamformers $\{\mathbf{w}^{\mathrm{comm}},\mathbf{w}^{\mathrm{sens}}\}$ 
to be reused in the next Stackelberg iteration.
\subsubsection{Belief Initialization and Scope}
\label{subsec:belief_init}
At the start of the adaptive phase, the ISAC BS initializes its posterior belief from the Bayesian–Stackelberg stage:
\begin{equation}
\mathbf{p}^{\mathrm{ADP}}_0=\mathrm{normalize}\!\big(P_{\mathrm{eve\_init}}\big)\in\Delta^{N_\phi-1}.
\end{equation}

Within each slot, Algorithm~\ref{alg:adaptive} refines the posterior belief $\mathbf{p}_t$, sensing weight $\gamma_t$, and beamformers $\{\mathbf{w}^{\mathrm{comm}},\mathbf{w}^{\mathrm{sens}}\}$, which are then used by Algorithm~\ref{alg:isac_BayStack}  for leader–follower role assignment and secrecy evaluation. Hence, Algorithm~\ref{alg:adaptive} optimizes where to sense and beamform, while Algorithm~\ref{alg:isac_BayStack} decides how to act through secure transmission or cooperative jamming.
\subsubsection{Resultant Security–Complexity Utility and Algorithm Coupling}
\label{subsec:final_utility}

After Algorithm~\ref{alg:isac_BayStack} executes, the ISAC BS (leader) obtains post-decision metrics:
\begin{equation}
\begin{aligned}
\{R_s^{(u)}(t)\}, \quad
\overline{R}_s(t) &= \frac{1}{|\mathcal{A}_t|} \sum_{u} R_s^{(u)}(t), \\
R_s^{\max}(t) &= \max_{u} R_s^{(u)}(t)
\end{aligned}
\end{equation}
$P_{\mathrm{out}}(t)=\tfrac{1}{|\mathcal{A}_t|}\!\sum_u \mathbf{1}\{R_s^{(u)}(t)<R_{\mathrm{th}}\}$.
The integrated leader–follower objective jointly rewards secrecy and efficiency:
\begin{equation}
\resizebox{\columnwidth}{!}{$
\begin{aligned}
U_t^{\mathrm{final}} =\ & \omega_{\max} R_s^{\max}(t)
+ \omega_{\mathrm{avg}} \overline{R}_s(t)
- \omega_O P_{\mathrm{out}}(t)
- \omega_H H(\mathbf{p}_t)
- \omega_C C_t \\
& - \lambda_O [R_{\mathrm{th}} - \overline{R}_s(t)]_+
- \lambda_H [H(\mathbf{p}_t) - H^\star]_+
\end{aligned}
$}
\label{eq:U_final}
\end{equation}
where $\omega_{\max},\omega_{\mathrm{avg}},\omega_O,\omega_H,\omega_C\!\ge\!0$ are design weights, and $(\lambda_O,\lambda_H)$ are dual variables updated in Algorithm~\ref{alg:isac_BayStack}. 
The interaction between the two algorithms forms a dual-loop ISAC control framework that ensures low-latency operation while adaptively enhancing physical-layer security under dynamic and uncertain conditions.

Hence, Algorithm~\ref{alg:adaptive} acts as the belief and beam refinement engine feeding improved posterior knowledge and sensing configuration into the Bayesian–Stackelberg leader–follower loop. This cross-coupled structure ensures convergence toward low-entropy, high-secrecy states with minimal computational overhead.
\begin{algorithm}[t]
\caption{Adaptive Sensing \& Beamforming (Per Slot)}
\label{alg:adaptive}
\begin{algorithmic}[1]
\Require Prior $\mathbf{p}_{t-1}$, kernel width $\sigma_t$, sensing map $\Gamma(\cdot)$, confidence threshold ${\rm conf}_{\rm th}$, protected angles $\mathcal{B}_t$ (from Alg.~\ref{alg:isac_BayStack})
\Ensure Posterior $\mathbf{p}_t$, sensing weight $\gamma_t$, comm/sensing beam weights $\{\mathbf{w}^{\mathrm{comm}},\mathbf{w}^{\mathrm{sens}}\}$, updated $\sigma_{t+1}$
\State \textbf{Belief prediction:} $\mathbf{p}^{\mathrm{pred}}_t\gets \mathbf{K}(\sigma_t)*\mathbf{p}_{t-1}$
\State \textbf{Confidence \& beam targeting:} $H_{\mathrm{tmp}}\!\gets\!H(\mathbf{p}^{\mathrm{pred}}_t)$, ${\rm conf}\!\gets\!1-\min(1,H_{\mathrm{tmp}}/\log_2 N_\phi)$; select top-$2$ MAP peaks (excluding $\mathcal{B}_t$) for sensing beams
\State \textbf{Adaptive likelihood:} $k_{\mathrm{eff}}\!\gets\!k_{\min}+(k_{\max}-k_{\min})\,{\rm conf}$; form normalized response $z(\phi)$ and update posterior: $\mathbf{p}_t\propto \mathbf{p}^{\mathrm{pred}}_t\odot z(\phi)^{k_{\mathrm{eff}}}$; early-stop if ${\rm conf}\ge{\rm conf}_{\rm th}$
\State \textbf{Beam design (low-complexity):} 
\emph{Sensing beam} $\mathbf{w}^{\mathrm{sens}}$: tapered focus on MAP peaks of $\mathbf{p}_t$ and \emph{Comm beam} $\mathbf{w}^{\mathrm{comm}}$: posterior-aligned, with notches on $\mathcal{B}_t$ (from Alg.~\ref{alg:isac_BayStack})
\State \textbf{Meta-kernel update:} $\sigma_{t+1}\!\gets\!\operatorname{clip}\big(\sigma_0[\,1+k_{\rm ex}\mathbf{1}\{H(\mathbf{p}_t)>H^\star\}-k_{\rm ey}\mathbf{1}\{H(\mathbf{p}_t)\le H^\star\}\,]\big)$
\State \textbf{Expose outputs to Alg.~\ref{alg:isac_BayStack}:} send $\mathbf{p}_t$, $\gamma_t$, and beam weights $\{\mathbf{w}^{\mathrm{comm}},\mathbf{w}^{\mathrm{sens}}\}$
\end{algorithmic}
\end{algorithm}
\section{Numerical evaluation}
\label{analysis}
\begin{figure}[h!]
\centering
\begin{minipage}[t]{0.48\columnwidth}
    \centering
    \includegraphics[width=\linewidth]{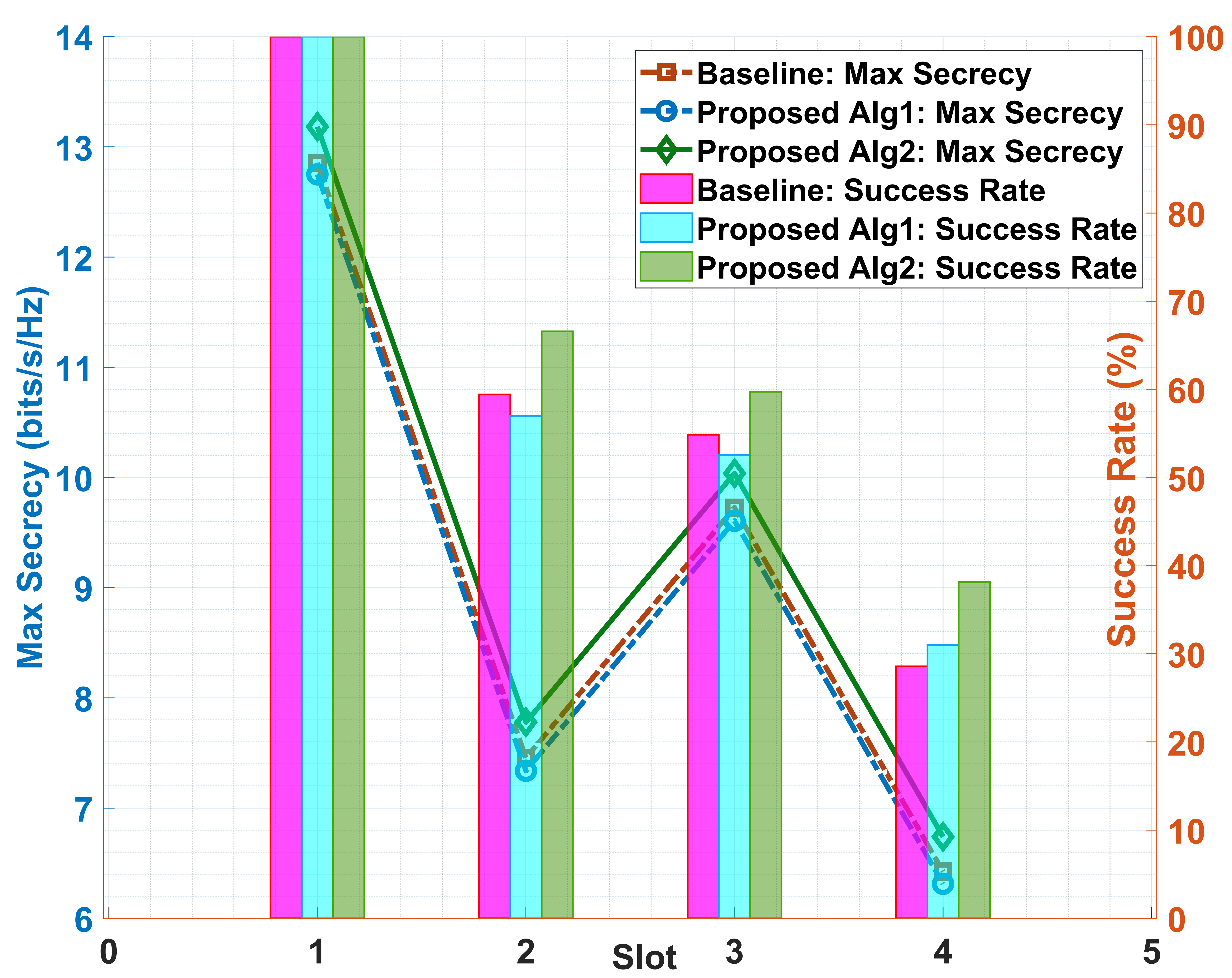}
    \caption{Comparison of maximum secrecy rate ($R_s$) and success rate (\%) across four slots over high-frequency bands from 28–410 GHz}
    \label{fig:secrecy_success}
\end{minipage}
\hfill
\begin{minipage}[t]{0.48\columnwidth}
    \centering
    \includegraphics[width=\linewidth]{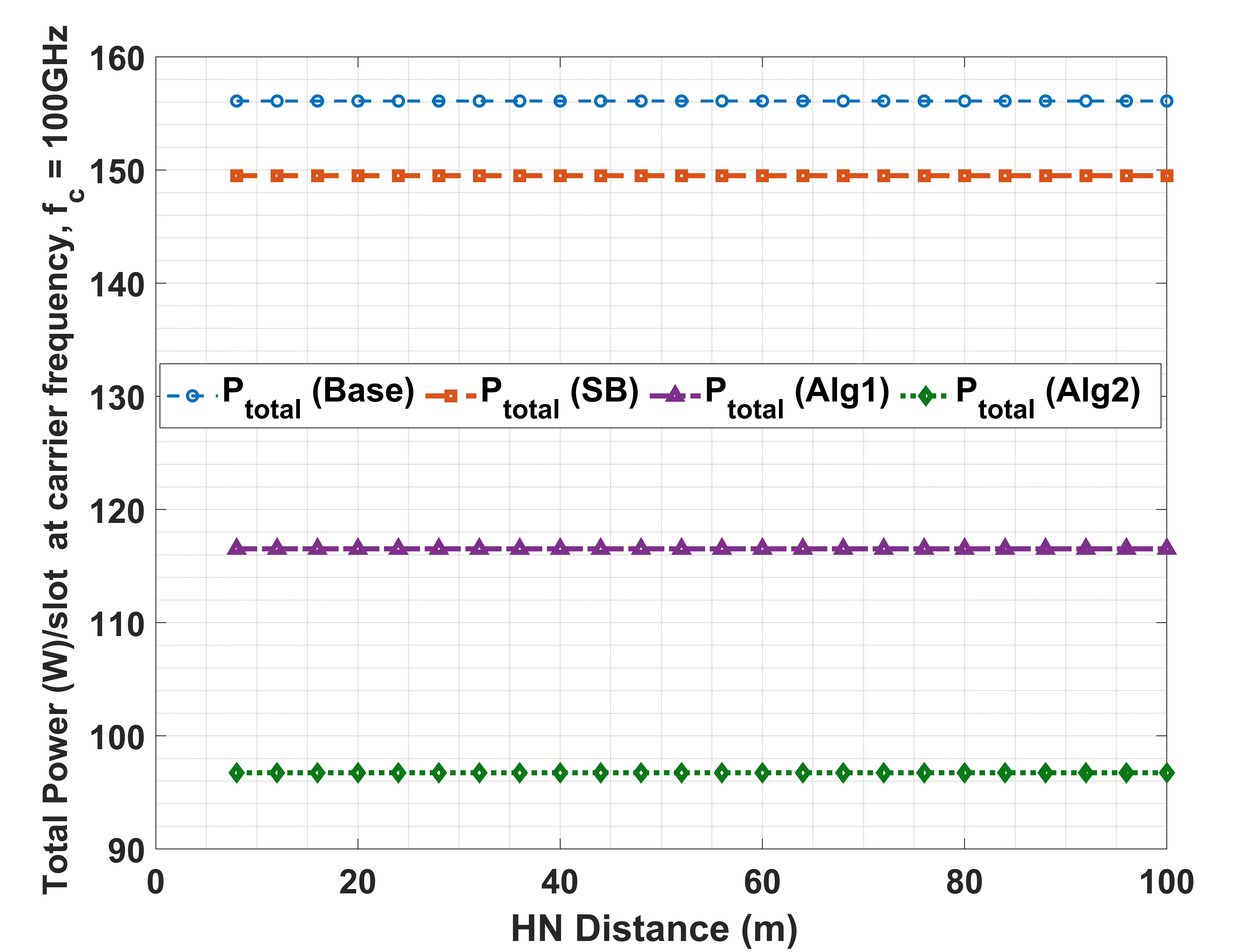}
    \caption{Total slot power versus HN distance at $f_c=100$\,GHz.}
    \label{fig:power_total}
\end{minipage}
\end{figure}

In this section~\ref{analysis}, comprehensive simulations were performed across carrier frequencies from $28$\,GHz to $410$\,GHz using the parameters listed in Table~\ref{tab:sim_params}. The ISAC base station (BS), equipped with $128$ antennas and $64$ RF chains, served $24$ hybrid nodes (HNs) within a $100$\,m coverage range. All simulations employed a $500$\,MHz bandwidth, $7$\,dB noise figure, maximum transmit power of $50.12$\,W, power amplifier efficiency $\eta_{\mathrm{PA}} = 0.38$, and a fixed secrecy threshold of $1$\,bit/s/Hz.
Firstly, Fig.~\ref{fig:secrecy_success} depicts the joint evolution of maximum secrecy and success rate across four slots. The baseline exhibits lower secrecy ($<10$\,bit/s/Hz) and higher outage (success rate $<40\%$), whereas Algorithm \ref{alg:adaptive} achieves the highest secrecy ($>13$\,bit/s/Hz) with success rates exceeding $95-98\%$. Algorithm\,1 provides intermediate gains while maintaining moderate computational cost. These results confirm the improvement of secrecy performance across various frequencies.
\begin{table}[h!]
\centering
\caption{Simulation Parameters}
\label{tab:sim_params}
\begin{tabular}{l l}
\hline
\textbf{Parameter} & \textbf{Value} \\
\hline
Carrier frequency, $f_c$ & 28–410\,GHz \\
Bandwidth (BW) & 500\,MHz \\
Noise figure (NF) & 7\,dB \\
Thermal noise $N_0$ & –174\,dBm/Hz \\
ISAC BS antenna arrays $N_T^{\mathrm{IBS}}$ & 128 \\
RF chains $N_{\mathrm{RF}}$ & 64 \\
HN antenna arrays $N_t^{\mathrm{UE}}, N_r^{\mathrm{UE}}$ & 16, 16 \\
Hybrid nodes (HN) & 24 \\
Streams per slot ($K$) & 6 \\
Initial TX power $P_{\mathrm{TX,cap}}$ & 50.12\,W \\
Power Amplifier, PA efficiency $\eta_{\mathrm{PA}}$ & 0.38 \\
Digital boost $\Delta_{\mathrm{BB}}$ & 20\,dB (+2\,dB Final) \\
Secrecy threshold $R_{\mathrm{th}}$ & 1.0\,bit/s/Hz \\
Beam notches (Algorithm 1) & (26\,dB,18°), (24\,dB,20°) \\
Beam notches (Algorithm 2) & (30\,dB,14°), (28\,dB,16°) \\
\hline
\end{tabular}
\end{table}
Secondly, Figs.~\ref{fig:utilities_alg1}–\ref{fig:utilities_alg2} depict the per-slot utility trends across various carrier frequencies. As frequency increases, utility decreases due to higher propagation losses; however, both proposed algorithms consistently surpass the baseline. At lower bands ($28$–$60$,GHz), they achieve strong secrecy gains, while in the THz range ($300$–$410$,GHz), Algorithm~2 maintains stable performance through optimized notch shaping. The subplots further show that utility degradation aligns with HN distance and path-loss variations.
\begin{figure*}[ht]
     \centering
    \begin{minipage}{0.48\textwidth}
         \centering
         \includegraphics[width=0.98\textwidth]{Wirelesscom_Figure/utilities_fc_alg01.png}
         \caption{Per-slot utilities of Baseline and Algorithm\ref{alg:isac_BayStack} across carrier frequencies}
         \label{fig:utilities_alg1}
     \end{minipage}
     \hfill
    \begin{minipage}{0.48\textwidth}
         \centering
         \includegraphics[width=0.98\textwidth]{Wirelesscom_Figure/utilities_fc_alg02.png}
         \caption{Per-slot utilities of Baseline and Algorithm\ref{alg:adaptive} across carrier frequencies}
        \label{fig:utilities_alg2}
     \end{minipage}
\end{figure*}
To further evaluate power optimization under low-complexity designs, Fig.~\ref{fig:power_total} presents the total per-slot power consumption at $100$,GHz. The baseline model exhibits the highest power usage ($\approx155$,W), followed by the Stackelberg–Bayes approach without ISAC integration ($150$,W), Algorithm \ref{alg:isac_BayStack} ($117$,W), and Algorithm \ref{alg:adaptive} ($95$,W). This progressive reduction in power consumption confirms that improvements in secrecy and reliability are achieved alongside lower power optimization.
\begin{table}[h!]
\centering
\caption{Total Runtime (s) vs Frequency for Two CPUs}
\label{tab:runtime_compact}
\begin{tabular}{|c|c|c|c|c|c|}
\hline
\textbf{GHz} & \textbf{i7-240H} & \textbf{Xeon E5} & \textbf{GHz} & \textbf{i7-240H} & \textbf{Xeon E5} \\
\hline
28  & 0.7233 & 1.0021 & 220 & 0.7775 & 0.9981 \\
60  & 0.7540 & 1.0545 & 300 & 0.6851 & 0.9794 \\
120 & 0.7984 & 1.0267 & 340 & 0.5952 & 0.9754 \\
     &        &        & 410 & 0.7616 & 0.9823 \\
\hline
\end{tabular}
\end{table}
Moreover, to verify the lightweight nature and computational efficiency of the developed algorithms, Table~\ref{tab:runtime_compact} summarizes the total runtime across different frequencies for two processing platforms. The Intel i7-240H processor achieves an average runtime below $0.8$,s per iteration, while the Xeon E5 processor maintains comparable stability at approximately $1.0$,s. Runtime remains nearly invariant with respect to frequency, reflecting the linear computational complexity of the belief-update mechanism and the early-stopping convergence design. Overall, the results at \ref{analysis} demonstrate that the proposed algorithms progressively enhance both secrecy capacity and success probability compared to the baseline. In particular, Algorithm~2 achieves up approximately 35\% greater secrecy performance, while maintaining negligible computational latency, confirming the practicality of the framework for real-time ISAC security in sub-THz and THz frequency bands.
\section{Conclusion and Future Studies}
\label{conclusion}
This work proposed a low complexity and robust security framework for near field ISAC systems across mmWave to THz frequencies. By combining Bayesian Stackelberg game control with adaptive hybrid node role switching, the framework achieves resilient physical layer security under dynamic and uncertain conditions. The numerical results verified consistent secrecy and reliability improvements with negligible runtime overhead. Future research will explore large scale distributed ISAC networks, AI driven belief updates, and RIS assisted extensions for enhanced scalability and real time deployment.
\bibliographystyle{IEEEtran}
\footnotesize
\bibliography{reference}
\end{document}